\documentclass[onecolumn,usenatbib]{mn2e}

\usepackage{graphicx}
\usepackage{rotfloat}
\usepackage{amssymb}
\usepackage{amsmath}
\usepackage{subfig}
\usepackage{comment}
\usepackage{multicol}

\begin{document}

\title[Early evolution of clumps]{Early evolution of clumps formed via gravitational instability in protoplanetary disks; precursors of 
Hot Jupiters?}

\author[M. Galvagni et L. Mayer]{M. Galvagni$^1$ and L. Mayer$^1$\\
  $^1$Institute of Theoretical Physics, University of Zurich,
  Winterthurerstr~190, 8057 Zurich, Switzerland}

\maketitle

\begin{abstract}

Although it is fairly established that Gravitational Instability (GI) should occur in the early phases of  the evolution of a protoplanetary disk, the fate of the clumps resulting from disk fragmentation and their role in planet formation is still unclear. In the present study we investigate semi-analytically their evolution following the contraction of a synthetic population of clumps with varied initial structure and orbits coupled with the surrounding disk and the central star. Our model is based on recently published state-of-the-art 3D collapse simulations of clumps with varied thermodynamics. Various evolutionary mechanisms are taken into account, and their effect is explored both individually and in combination with others: migration and tidal disruption, mass accretion, gap opening and disk viscosity. It is found that, in general, at least 50\% of the initial clumps survive tides, leaving behind potential gas giant progenitors after $\sim 10^5$ yr of evolution in the disk. The rest might be either disrupted or produce super-Earths and other low mass planets provided that a solid core can be assembled on a sufficiently short timescale, a possibility that we do not address in this paper. Extrapolating to million year timescales, all our surviving protoplanets would lead to close-in gas giants. This outcome might in part reflect the limitations of the migration model adopted, and is reminiscent of the analogous result found in core-accretion models in absence of fine-tuning of the migration rate. Yet it suggests that a significant fraction of the clumps formed by gravitational instability could be the precursors of Hot Jupiters.

\end{abstract}


\begin{multicols}{2}

\section{Introduction}

The study of planet formation has been boosted in the last decade due to unprecedented observational campaigns of extrasolar planets enabled by new telescopes and 
techniques \citep[see][]{2012arXiv1205.2273W,2011ApJ...728..117B,2011ESS.....2.0103S,2010ApJ...719..497L}. The large diversity of physical characteristics that these objects show \citep{ma2013} leads to the idea that there must be more then one mechanism to generate them. Indeed, they are very different for mass, composition (from rocky planets to gas giants), radii and position in the disk (from a few fractions up to hundreds of au from the central star).

The state of art main formation scenarios are Core Accretion (hereafter CA) and Gravitational Instability (hereafter GI) \citep[for a review see][]{armitage}. While CA is generally recognized as the mechanism by which most planets should form, and it is by construction meant to form gas giants as well as rocky planets, GI has received revived interest with the discovery of gas giants at large distances from their parent stars (a $>$ 30 au) since this is the region where disks are likely undergoing fragmentation in the early stages unless they are stabilized by strong irradiation \citep{rafikov2009,boley2009,zhu2012}. On the theoretical side,  GI, which has been traditionally restricted to explain gas giants \citep{boss1997,mayer2002}, has been developed in a new direction in the last few years as it has been recognized that Tidal Downsizing coupled with accretion of solids and core formation within gas clumps can in principle lead to Super-Earts and other rocky planets \citep{boley2010,nayakshin2010}. Furthermore, recent work has shown that radial migration plays an important role in GI as it is already known to play in CA \citep{baruteau}, possibly leading to planets at distances much lower than those of their formation site. It is therefore clear that the fate of clumps produced by GI depends on several mechanisms, many of which the same that are also crucial in CA. While this adds complexity, it is also the sign that GI has now become a much more mature theory, within which predictions can now be made beyond the short timescales probed by simulations \citep{zhu2012} by combining analytical calculations of several processes, as it has been done in CA for quite a few years \citep{alibert2005}. Among other processes that have been recentely considered in GI studies, there are the sedimentation of dust and core formation \citep{boley2010,nayakshin2010,forgan} This potentially allows to start making predictions that can be verified or falsified by observations, as it has been done now for a few years with CA, and finally allows to compare both formation theories on the same ground.

Indeed, although the concept of GI as a giant planet formation model has been around for a long time \citep{kuiper}, studies of the long term dynamical evolution of clumps formed in GI are just now beginning to appear \citep{zhu2012}. This isn't the case of CA, where studies of population synthesis models \citep{mordasini2009} have been proposed over the last years, making it possible to produce statistical expectation of the characteristics of extrasolar planets formed via this mechanism.

While fully radiative 3D simulations are too expensive to allow carrying out studies of clump evolution on long timescales \citep{boley2010}, simpler 2D simulations with phenomenological cooling pescriptions \citep{vorobyov2013} have recently been used to study the likelihood that fragments are the progenitors of the giant planets and brown dwarfs that are detected at tens of au from the hosting star \citep{marois2008,kalas2008}. These works find that most of the clumps migrate inward rapidly and are destroyed by tides in the inner region of the disk before they can become full fledged planets, leading to the conclusion that succesfull planet formation by GI is a rare occurrence in general. However, these simulations have low resolution (a few AU with their grid size), which likely leads to artificial clump disruption by tides at small radii, an effect that is known to have plagued simulations of self-gravitating collapse with non-adaptive grid techniques in other fields of astrophysics, such as star formation and cosmological structure formation \citep[for effects of resolution on disk fragmentation in various numerical techniques]{durisen2007,mayer2008}. Futhermore, no existing numerical simulation of self-gravitating protoplanetary disks by either grid-based or SPH codes, has enough resolution ro resolve the internal structure of the clumps and allow studying properly their collapse. Clump collapse is crucial since it will determine the response of the clumps to migration and tides by affecting its density, mass and temperature, as shown by analytical studies that focus on this process \citep{helled2011,vazan2012}.

In order to be able to make prediction of the characteristics that a population of extrasolar planets formed via GI would present, more accurate studies of the very early stage of clump formation and evolution are needed. Indeed, the main question regarding GI is: are the clumps that form going to survive the interaction with the disk and hosting star, or will they be disrupted? 

Fortunately, the first high-resolution 3D fully hydrodynamical simulations of clump collapse have been performed \citep[][and Galvagni et al. in prep.]{galvagni}, adding a new important step that goes in the direction of answering these questions properly. The results of the published collape simulations constitute the backbone of this paper.

As a first approximation, we can divide the lifetime of a clump into two parts: pre-dissociation and post-dissociation phase. It is indeed known that, while a clump contracts, it will eventually reach inner temperature and density high enough to dissociate molecular hydrogen \citep{masunaga1998}. During this phase, the gas behaves as if it were almost isothermal, with an effective adiabatic index of $1.1$. This is due to the fact that a fraction of the thermal energy increased by gravitational collapse is used to break the molecular bound and doesn't lead to an increase of temperature, and as a result the pressure support is reduced. This leads to a fast collapse that shrinks the clump into a more compact object. Once the clump has undergone this process, it can be safety assumed to be so compact that it would resist interaction with the disk and hosting star, being then a real protoplanet.

Due to its much longer timescale phase the key phase is thus the first phase of the clump life: if it is able to reach the dissociation of hydrogen (called second core collapse, hereafter SCC) without being priorly disrupted, then we can assume it is a real protoplanet. 

\cite{nayakshin2012,zhu2012,forgan} recently presented first attempts towards a population synthesis model for GI, by coupling the evolution of clumps during the first collapse phase with  their interaction in the disk. 

However, none of these works relies on accurate clump collapse simulations, rather it assumes a timescale for the collapse in the first phase or it computes that while assuming quasi-static collapse based on the notion that the dynamical time is always much shorter than the cooling time. However, detailed studies of clump collapse assuming near hydrostatic equilibrium at all times obtain clump collapse timescales that are up to 2 orders of magnitude longer relative to those found in the 3D hydro simulations. Since the relation between the timescale of the various processes involved is ultimately what will decide the fate of the clumps \citep{boley2011,nayakshin2010,zhu2012}, starting from a self-consistent model of internal clump evolution becomes a pivotal factor.

The aim of the study herein presented, is to address the following question: how many of the clumps formed via GI in a standard circumstellar disk will survive to SCC? And what will be their characteristics? The answer will give an estimate of how luckily is GI to be a valid mechanism for forming planets. Nevertheless, this work does not represent a synthesis population model for GI, as the final characteristics of the population of planets formed via this mechanism still depend on the evolution that the clump undergoes from SCC on. It is indeed expected that the evolution of clumps after SCC will be dominated by scattering, migration and dynamical interactions between clumps and the hosting star. This late evolutionary phase is therefore crucial in explaining the characteristics of observed planets, such as the misalignment of Hot Jupiters \citep{ma2013}. Indeed, it has been proposed that dynamical interactions alone can be the origin of Hot Jupiters \citep{chatterjee2008}. A prediction of these quantities and features is therefore beyond the scope of this work, and our results should be taken only as an estimate of the final position and mass of the planets when SCC happens, not of the final planet population.

The paper is organized as follow: the next section presents the methods, with detailed explanation of the implementation of the different physical mechanisms taken into account. Section \ref{sec:res} presents the results of our simulations, which are discussed in section \ref{sec:dis}. The conclusions are given in section \ref{sec:con}.

\section{Methods}

We consider the evolution of a set of clumps formed via GI inside a disk, from the fragmentation of spiral arms to SCC. Since our focus here is to study the fate of clumps {\it provided that they form by GI}, rather than the conditions to form clumps by GI in disks, we will not study the fragmentation phase of a Toomre-unstable disk as done in eg. \cite{zhu2012,forgan}. Instead, we  will assume that clumps are already present and study their evolution under the combined action of all the key mechanisms: collapse, migration, mass accretion, and tidal effects. As in all the semi-analytical works on the subject so far, we will not include the effect of the dynamical interaction between different clumps, which is known to take place in 3D disk simulations, leading occasionally to clump-clump merging and rendering their orbital dynamics more stochastic than expected if only inward radial migration takes place \citep{boley2010,zhu2012}. Therefore, in our model we are essentially considering the simple situation in which there is one clump per disk and, therefore, by generating a population of clumps as we will do, we are following a population of protoplanetary disks in which fragmentation has taken place. What fraction of the overall disk population the latter population represents is beyond the scope of this paper. Considering only one clump per time is a simplification and is therefore a first step towards the development of a complete population synthesis model for GI. 3D simulations, indeed, show that usually the formation of a first clump in a GI unstable disc triggers subsequent fragmentation of spiral modes, that leads to the formation of 3-4 clumps per time\citep{kratter2006,meru2013b,forgan2013b}. However, 2D simulations, which are less accurate than 3D simulations but can probe a larger parameter space, show that in most cases clump-clump interactions are not the dynamical dominant process \citep{zhu2012}. Therefore we do not expect this simplification to have a major impact on the final results.

We generate a population of clumps in a random way (see section \ref{sec:Mdis}) and evolve each of them independently. Three different scenarios are explored (see table \ref{tab:sim}). In all the scenarios, clump contraction, migration and tidal disruption are taken into account. In scenario B, mass accretion from the disk onto the clump is added. Scenario C implements also a gap opening criteria. For each scenario, different sets of simulations are run, changing the initial conditions and exploring different gap opening implementations. 

This study concentrates on the formation of gas giant planets; we therefore neglect as a first approximation grains coagulation and core formation. Due to this approximation, we are not able to capture the formation of rocky planets through tidal downsizing.

The disk model and initial conditions are presented in \ref{sec:Mdis}; all the phenomena implemented in the different scenarios are presented from \ref{sec:Mcon} to \ref{sec:gap}.

\subsection{Disc model and Initial Conditions} \label{sec:Mdis}

\end{multicols}

\begin{figure}
\centering
\subfloat{\includegraphics[width = 0.30\textwidth]{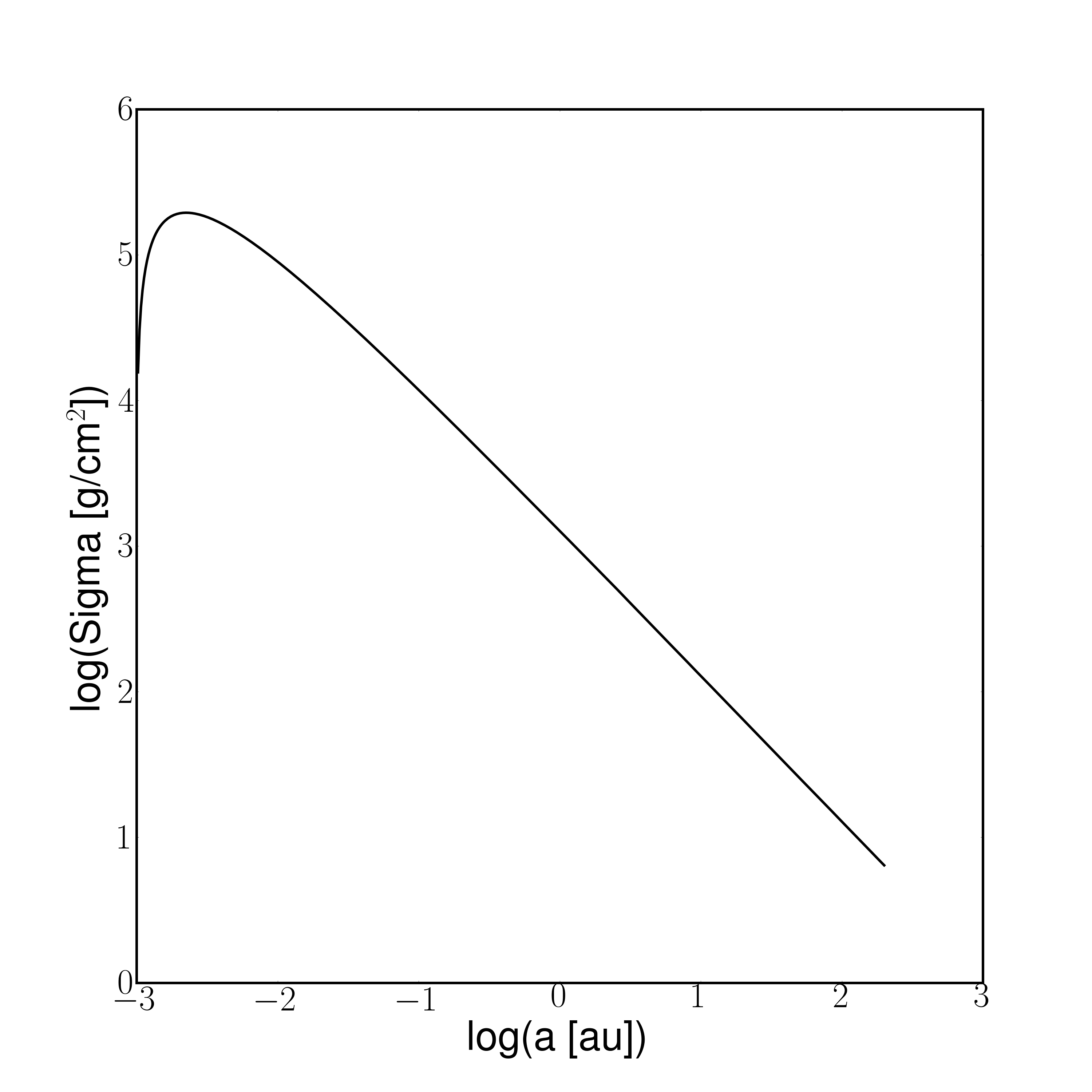}}
\subfloat{\includegraphics[width = 0.30\textwidth]{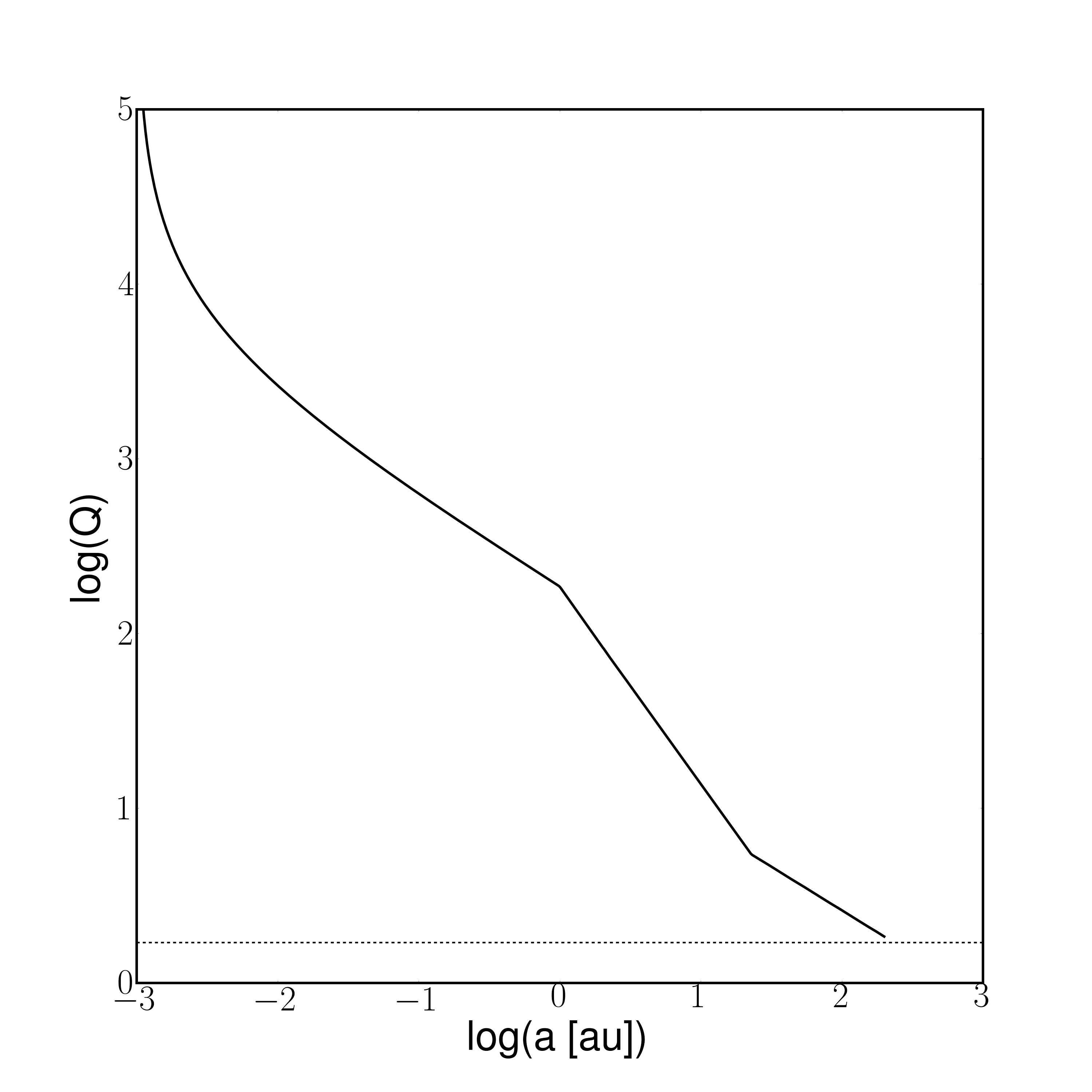}}
\caption{On the left, surface density profile of the circumstellar disk. On the right, Toomre parameter profile. The horizontal dashed line corresponds to the critical fragmentation value $Q_{\mbox{cr}} = 1.7$. Scales are logaritmic.}
\label{fig:disk}
\end{figure}

\begin{table}
\caption{Physical mechanisms implemented in the different scenarios.}
\begin{center}
\begin{tabular}{c||c||c||c||c||}
\hline
 & Migration+Tidal Disruption & Mass Accretion & Gap opening & Mass Accretion during gap \\
 \hline
 Set A & Yes & No & No & No\\ 
 Set B & Yes & Yes & No & No\\
 Set C & Yes & Yes & Yes & Yes \\
 Set C\_m0 & Yes & Yes & Yes & No \\
\hline
\end{tabular}
\end{center}
\label{tab:sim}
\end{table}

\begin{multicols}{2}

In a protoplanetary disks self-gravitating clumps can form from the fragmentation of spiral arms in Toomre unstable disks \citep{boss1997,mayer2002}.  Recently a new mechanism has been proposed for the formation of these objects \citep{hopkins2013} based on turbulence-induced fragmentation in self-gravitating disks occurring even when the disk is Toomre stable. We will not consider this case, as it is a less likely occurence and requires further scrutinity. The focus is then on clumps which form from the fragmentation of spiral arms when $Q_{min} < 1.4$ \citep{durisen2007}, as found by a wide range of calculations. We also assume that a disks self-regulates to a marginally stable steady state immediately after fragmentation ($Q_{min} = 1.7$, see disk Q profiles in Figure \ref{fig:disk}), neglecting the possibility of recurrent fragmentation epochs sustained by either disk mass loading or sudden opacity changes that shorten the cooling time, since this would require following disk evolution \citep{vorobyov2009a}. We emphasize once again that our model is intentionally simple and errs on the side more effects that would favour clump formation and survival rather than the opposite (see Summary/Discussion section on caveats).

Our starting point is thus the disk configuration soon after fragmentation. We generate two different sets of initial conditions. In each case, the only quantities we need to specify for the clumps are: initial semimajor axis $a$, mass clump $M_{\mbox{clp}}$ and mass radius $r$. 

Fragmentation is more likely in the  early stage of the disk lifetime, when the disk is still quite massive due to gas accretion from the environment and the star is still growing significantly in mass via accretion, as in Class 1-2 phases \citep{machida2010,eisner2012}. We therefore assume that the star mass is $M_{\mbox{star}} = 0.6 M_{\odot}$ while the disk mass is $30 \% M_{\mbox{star}} = 0.18 M_{\odot}$. According to the results in \cite{lodato2004,forgan2011a}, this choice of disk mass allows to safely assume the angular momentum transport to be local and therefore the viscosity of the disk can be parametrized via the dimensionless $\alpha$ parameter \citep{shakura1973}.
The observational dissipation timescale for a circumstellar disk is of the order of Myr \citep{haisch2001}, while the typical timescales for the evolution of clumps formed via GI is of the order of $10^5$ yr (collapsing and migrating timescales, see the next sections). We can therefore safely neglect the time evolution of the disk. Due to this simplification, we can adopt a steady state disk as our background disk model, whose  surface density 
profile $\Sigma$ is given by the numerical solution of the equation \begin{equation}
\frac{\partial }{\partial a} \left[ a^{1/2} \frac{\partial  }{\partial a} \left( \nu \Sigma a^{1/2} \right)\right] = 0
\end{equation}
as implied by the diffusion equation for a thin viscous disk in \cite{Lynden} when assuming $\partial \Sigma / \partial t = 0$. Figure \ref{fig:disk} 
show the surface density and the Toomre parameter profiles obtained.

In the above equation $\nu = \nu_0 a$ is the viscosity of the gas. We assume that the radius of the disk is $150$ au, as found in simulations. By fixing the total mass of the disk the ratio $K/\nu_0$ is determined, where $K$ is a constant of integration. Hence the viscosity remains a free parameter if only the mass of the disk is specified.

In both sets the initial semimajor axis $a$ is assumed to be in the region where the Toomre parameter is near the minimum, where it is most likely that clumps have been formed (while the disk profile adjusts as the disk self-regulates no large radial excursions are seen in the location of the minimum Q as long as the physical conditions in the disk do not change, see eg Mayer et al. 2004). The determination of the region where Q is minimum is affected by parameters such as the star and disk mass and the disk viscosity. At the same time, it is unaffected by the disk radius. This parameter indeed does not play a major role in determining the disk profile, as the value of $\Sigma$ in the outer part of the disk is small, and therefore a large change in the value of the disk radius leads to a minor redistribution of the mass and consequently to a small change in the Toomre profile. The only case where the disk radius could play a major role, would be considering a compact disk, where typically $a<60$ au. However, this is not the case explored in the present work, as GI would not be possible in this environment.
As it can be seen in figure \ref{fig:disk}, the latter region is between $80$ au and $120$ au.
In the first set (hereafter IC) the initial mass $M_{\mbox{clp}}$ and radius $r$ of the clumps are taken to match typical values found in GI simulations \citep{boss2011,galvagni}: $r$ is taken in $1.0-6.0$ au while $M_{\mbox{clp}}$ in the range $0.5 - 5.0 M_J$. The mass range is consistent with \cite{boley2010}, where it has been found that the typical mass of a fragment is nearly an order of magnitude lower than the local Toomre mass due to effects in the nonlinear regime of gravitational instability. 
A second set of more massive initial condition (hereafter ICM) is generated. In ICM the clump mass is assumed to
 be in the range $5 - 12 M_J$ and the clump radius has been rescaled to $2-12$ au in order to have a similar initial density in the two sets of initial condition. This second more massive set is meant to include, in a very first approximation form, the possibility of multiple fragmentation followed by merges between the clumps. It is indeed still unclear if the outcome of GI is usually single or multiple clump formation. In the latter case, merges between the clumps are expected, as they are massive (gravitational focusing) and they form in a relatively small region. Moreover, this second set of more massive clumps is consistent with the initial conditions assumed in \cite{vorobyov2013} and \cite{forgan}, or also with the masses of clumps formed in the earliest phases of protostellar disk evolution soon in the first few $10^4$ years following the collapse of the molecular cloud core \citep{hayfield2011} , making comparisons with other works more feasible.

\subsection{Clump contraction and migration} \label{sec:Mcon}

\end{multicols}

\begin{table}\footnotesize
\centering
\begin{tabular}{|c|c|c|}
\hline
\includegraphics[width = 0.25\textwidth]{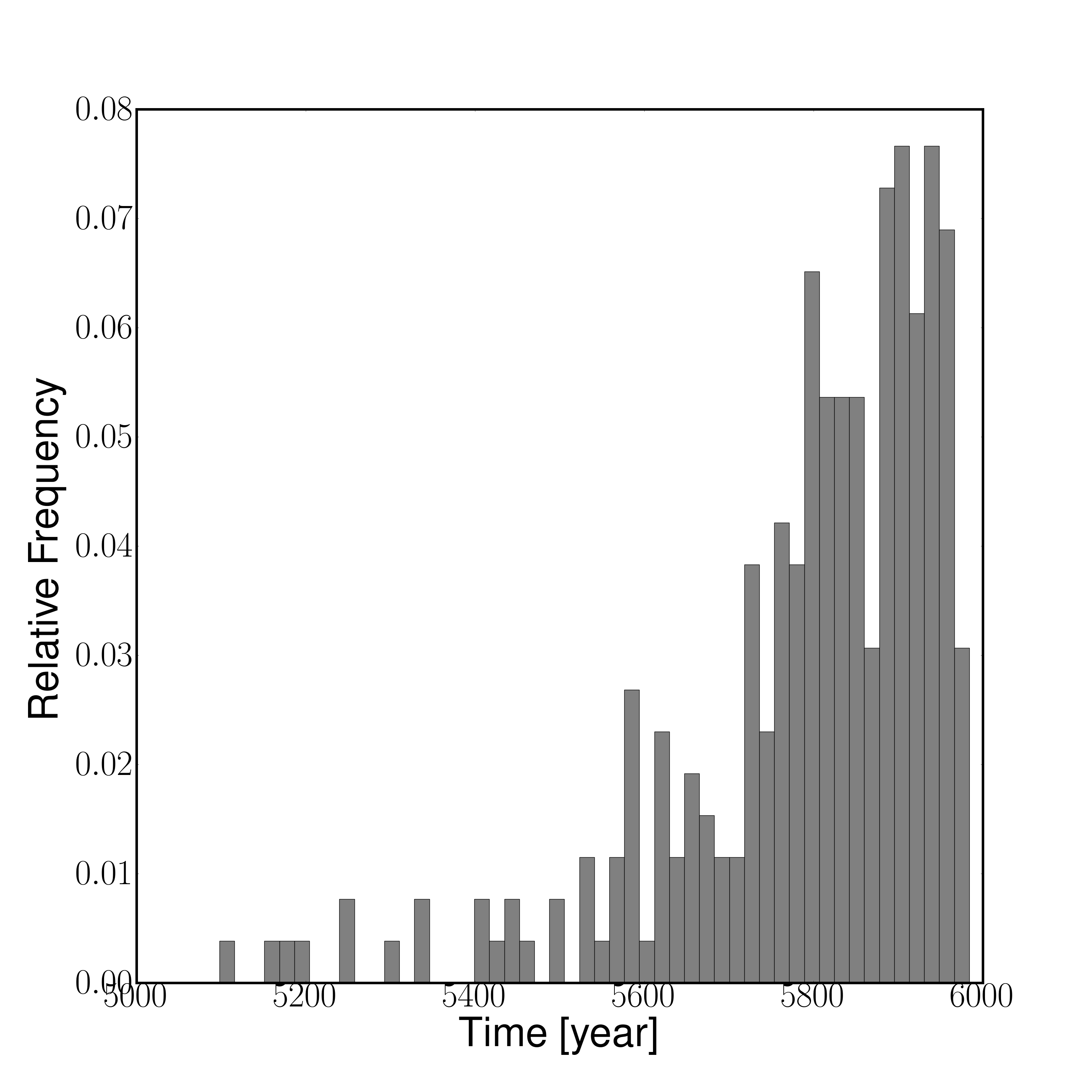} &
\includegraphics[width = 0.25\textwidth]{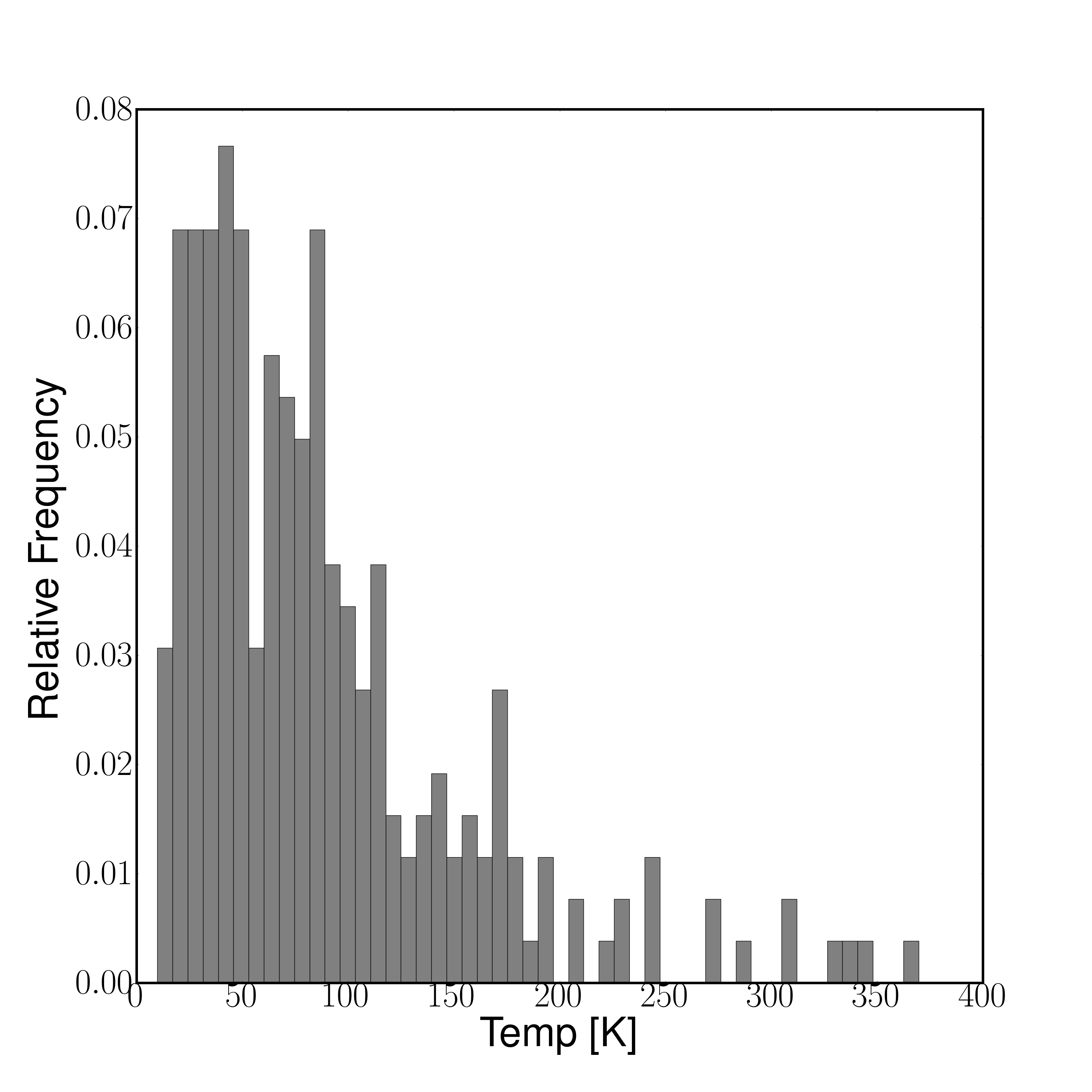} &
\includegraphics[width = 0.25\textwidth]{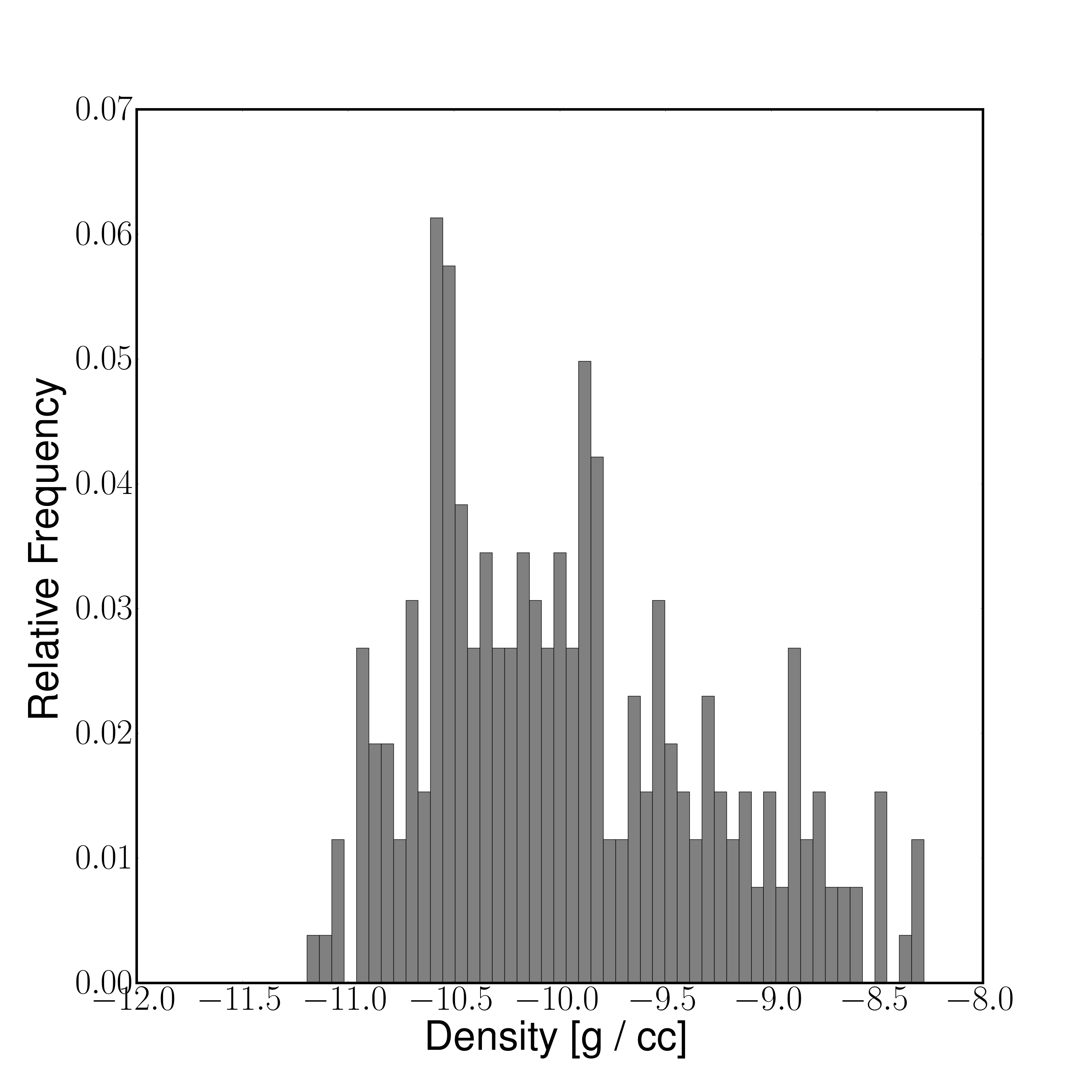} \\ 
\hline
\includegraphics[width = 0.25\textwidth]{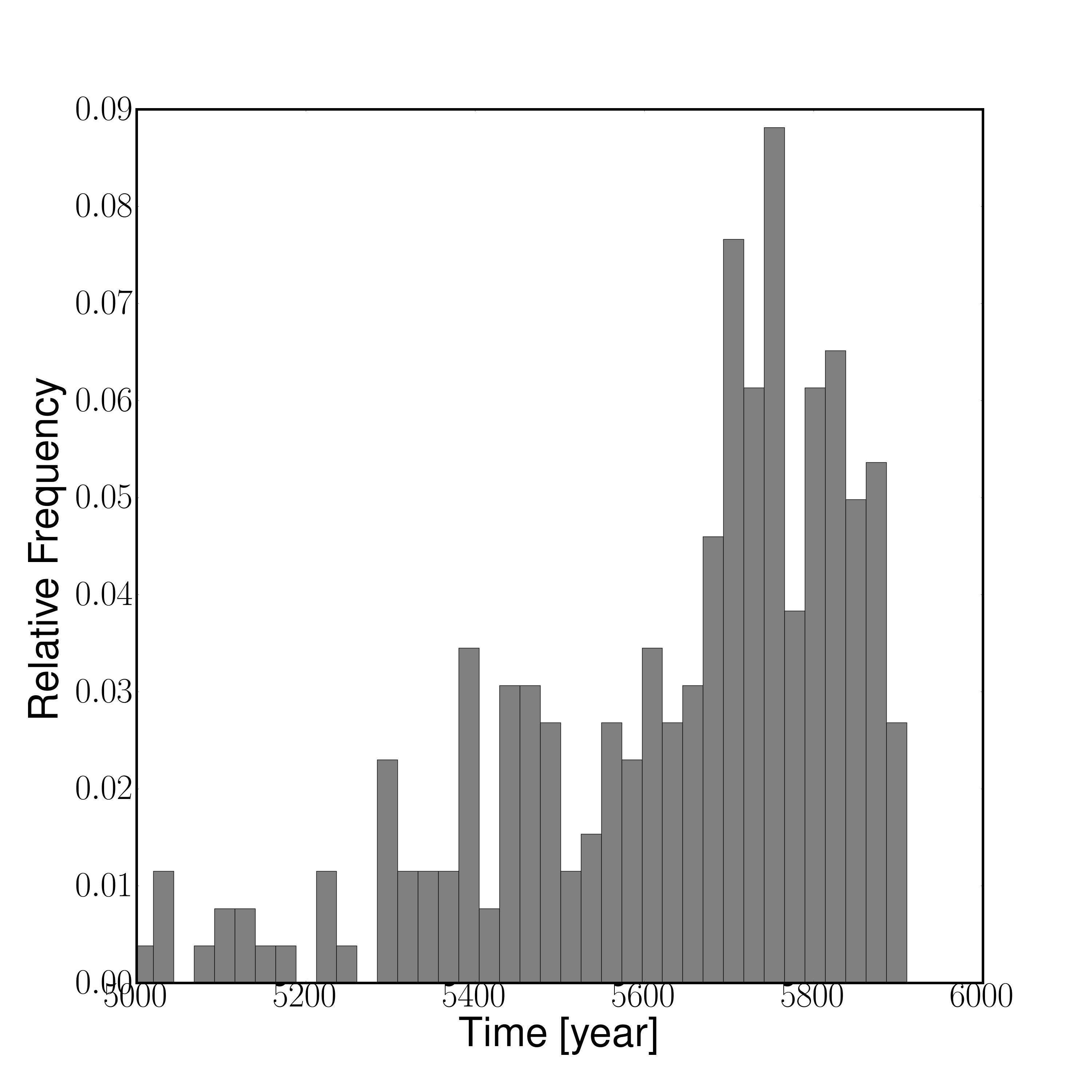} &
\includegraphics[width = 0.25\textwidth]{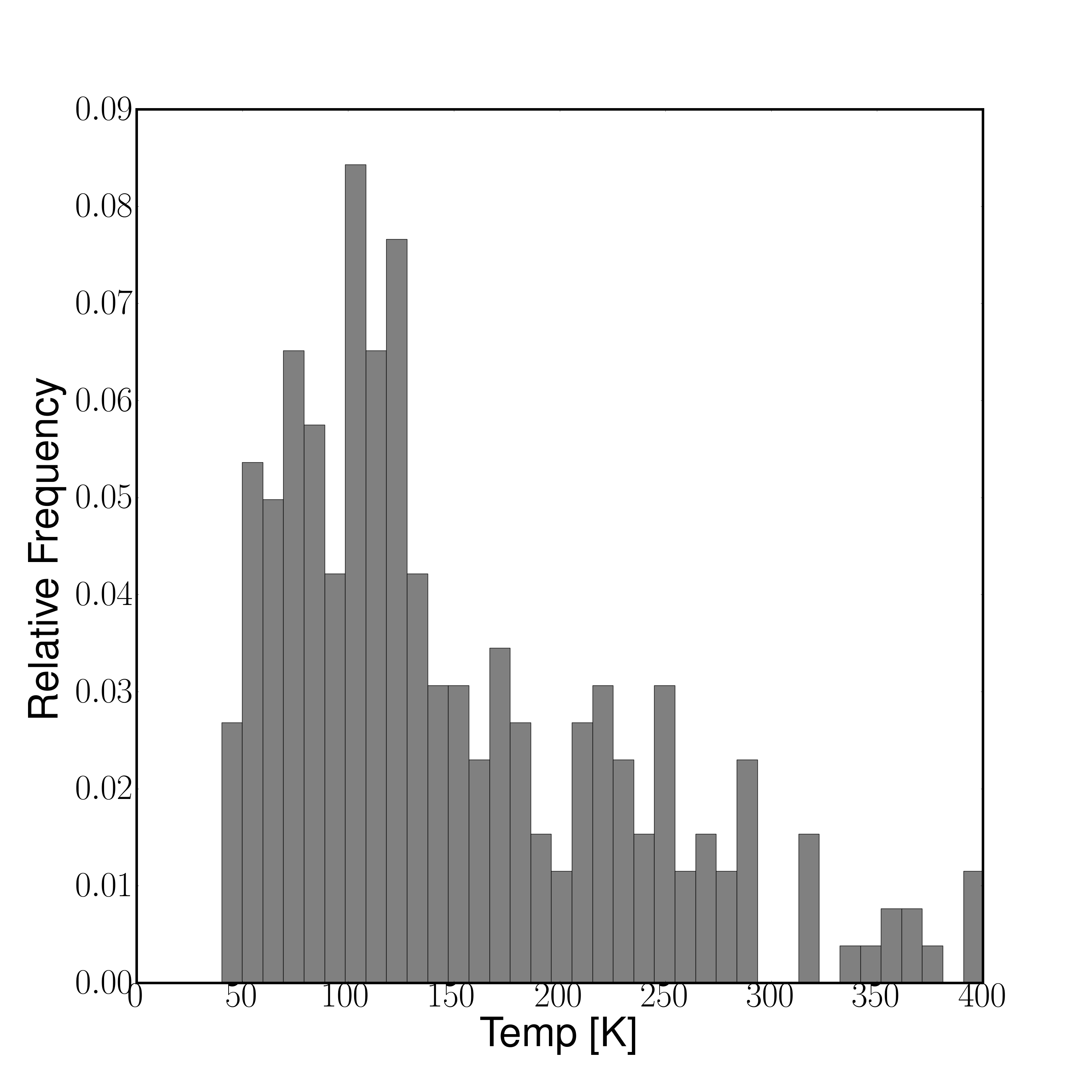} &
\includegraphics[width = 0.25\textwidth]{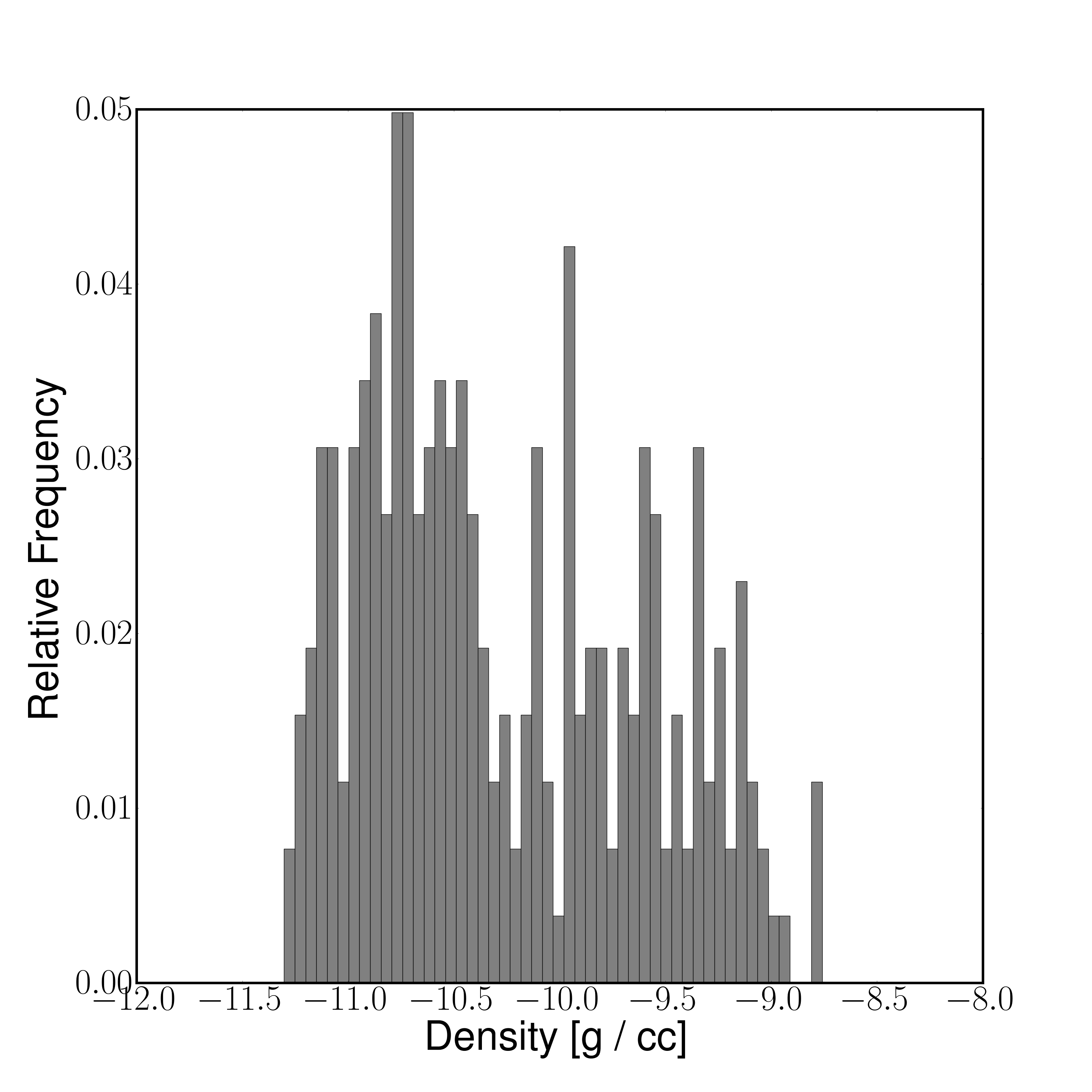} \\ 
\end{tabular}
\caption{Histograms of the initial conditions for the clumps in the set IC (on the top) and ICM (on the bottom). For each line: on the left, histogram of the contracting timescale for the clumps evolved; in the middle and on the right, respectively, histogram of the inner temperature and density (in log scale).}
\label{tab:ctime}
\end{table}

\begin{multicols}{2}

The fate of the clumps is determined from the competition between contraction and tidal disruption due to the presence of the central star, modulated by mass accretion and
radial migration which affect contraction and disruption. 
The contraction/collapse timescale is assumed to be the time to reach SCC, as it is known that when a clump undergoes second core collapse its size shrinks and its density increases by order of magnitudes \citep{masunaga1998}, making it safe from tidal disruptions and therefore a real protoplanet. The tidal disruption timescale on the other hand will depend on the migration timescale of the clump, as it can be disrupted only if it gets close enough to the star that its radius becomes large enough that the outer part is no more gravitationally bounded to the clump.

In this work the contraction timescale to reach SCC is assumed to be the time to have the dissociation parameter\footnote{The dissociation parameter is defined as the ratio between protons in the atomic form of hydrogen over the total number of protons in the gas.} in the inner part of the clump of $1$ \% \footnote{SCC is a non linear effect, with a strong back feeding component. Indeed, when dissociation of hydrogen starts, the fast collapse increases locally the gas temperature, triggering dissociation in gas parts where it was still not active. Dissociating parameter of $1$ \% is therefore a safe assumption to consider all the clump in SCC phase.}. We use the results presented in \cite{galvagni} to determine the contracting timescale for each clump. \cite{galvagni} presents 3D high-resolution study of the collapse of a realistic clump formed via GI, taking into account also its initial rotation and non axisymmetric state. The timescales therein presented are therefore the more realistic ones currently available. This work shows that when a clump forms from the fragmentation of a spiral mode due to GI, it undergoes a quite fast phase (of the order of a few dynamical times) where the initial asymmetries lead to a redistribution of the angular momentum. After this phase, the clump becomes spherical and its contraction becomes constant, so that it becomes possible to extrapolate the evolution of the inner density and temperature, and therefore the contracting timescale. 

Due to the presence of this first unstable phase, the clumps we generate as initial conditions are not the clumps that are formed just after fragmentation of the spiral mode, but a few dynamical times after that. In this way, it is possible to use in a confident way the extrapolation results obtained in the previous paper. Postponing the initial evolution time of the clumps leads to an uncertainty in the position of the clump in the disk, as it still undergoes migration in the very first phase we are neglecting. Nevertheless, as this neglected phase is fast, and as there already is an uncertainty on the initial semimajor position of the clump, we can safety assume that this procedure doesn't affect our results. The main consequence of neglecting this initial phase is that the initial temperature and density of the clump have larger values that the ones usually observed in simulations for clumps formed into spirals.

In order to perform a statistical study of the fate of clumps generate via GI, we need to generate a large population of realistic initial clumps. The initial values of density and temperature are calculated by rescaling the clump simulated in \cite{galvagni} to fit the mass and radius of the new clump:
\begin{eqnarray}
\rho' &=& \rho_0 \frac{M'}{M_0} \left( \frac{r_0}{r'} \right)^3 \\
T' &=& T_0 \frac{M'}{M_0} \frac{r_0}{r'}
\end{eqnarray}
with $r_0$, $M_0$, $\rho_0$ and $T_0$ radius, mass, inner density and temperature of the clump in \cite{galvagni} and $r'$, $M'$, $\rho'$ and $T'$ values for the random generated clump. The first rescaling comes from the definition of density. The second rescaling comes from assuming that the ratio between gravitational and thermal energy stays constant between different clumps. Figure \ref{tab:ctime} presents the distribution of contracting timescales, initial inner temperature and density for the two set of clumps generated in this work (IC and ICM). 

The derived collapsing timescale depends on the initial rotational status of the clump, and how it redistributes the internal angular momentum. Nevertheless, we can assume that the derived collapsing timescales are a safe overestimate. Indeed, the initial ratio between rotational and gravitational energy is $T/|W| = 0.15$ for the reference clump in \citep{galvagni}, but quickly gets to higher values during the first collapsing phase. Therefore, if the initial value were lower, it would mean that the clump is a very slow rotator, and the collapsing timescale would decrease. On the other hand, if the initial value of $T/|W|$ were larger, the strong rotational component would lead to the formation of a bar instability, which is a very effective structure for angular momentum redistribution. Therefore, the angular momentum of the inner part of the clump would be transferred outward, leading also in this case to a decrease of the collapsing timescale. 

Once the clump forms in the outer part of a disk, it starts migrating due to the interaction with the disk. The migration characteristics are strictly related to the local disk properties; it is usually inward, although it could be outward in some particular cases \citep{bate2003,nayakshin2012}. In this work, we assume that the migration is inward, following the result presented in \cite{baruteau}. Equation \ref{eq:tau} describes the migration timescale for a clump in a similar mass and radius range as the one we study:
\begin{equation} \label{eq:tau}
\tau = T_{\mbox{orb}} 5.6 (3.8-\sigma)^{-1} \gamma Q \left( \frac{q}{h^3} \right)^{-1} \left( \frac{h}{0.1} \right)^{-2}
\end{equation}
with $\sigma=1.3$ power of the density surface of the disk, $\gamma=5/3$ adiabatic index, $Q$ local Toomre parameter, calculated from the disk profile, $q$ ratio between clump and star mass, $h$ disk aspect ratio. The inward migration is stopped when the clump reaches $0.01$ au, as at that distance the magneto field coming from the star prevents it to migrate even further in. The value for the inner disk edge of $0.01$ au is chosen in the typical range for magnetospheric boundaries \citep{armitage}. Assuming that the clump stops when it reaches this value is a simplification, as mass accretion onto the star through the magnetic field lines is still possible in principle. Nevertheless, we assume that this process do not significantly affect the clump mass, as the Alfven velocity associated to a typical T Tauri star magnetic field of 1 kG \citep{armitage} is 3 orders of magnitudes lower that the escape velocity from the clump itself. We also impose a zero mass accretion onto the clump once it has reached the inner disk edge, in order to take into account the presence of competitive mass accretion onto the star through the magnetic field lines. As at the state of the art there is no precise knowledge of how this process happens, and therefore a correct treatment of it is out of the reach, the authors feel that our treatment is a reasonable simplification of the process. The migration timescale has been derived for clumps whose mass is in the range $q = 10^{-4} - 5 \times 10^{-3}$. According to the stellar mass value used in our simulations, these value correspond to a clump mass in the range $M_{\mbox{cl}} = 0.1 M_J - 5 M_J$. This doesn't mean that for values outside this range the migration formula doesn't apply, but that it has not been tested. In our semi analytical model, this mass range covers the majority of the studied clumps. Indeed, those whose mass is reduced to a value lower than $0.1 M_J$ by tidal stripping do not survive the stripping phase. As this phase is fast compared to migration timescales ($10^2$ years), a modification in the migration formula would not play any role in the final result. For clumps whose mass is higher than $5 M_J$, on the other hand, we still assume formula \ref{eq:tau} as we do not identify any physical reason why it should not hold in this regime as well, despite not having been directly tested.

The contracting and migration timescales are both of the order of $10^4$ years. This means, they have comparable values and therefore we need to follow the evolution of each clump in the disk in order to predict its fate. 

\subsection{Tidal disruption}

While the clump moves inward, the outer part feels the gravitational interaction from the hosting star. If the clump gets close to it faster then it contracts, the outer part is stripped away. The distance at which this happens is usually assumed to be the distance at which Hill radius $R_h = a \times (M_{\mbox{cl}}/(3M_{\mbox{star}}))^{1/3}$ and clump radius are equal. We assume that the outer part of the clump is disrupted when its radius equal one third of the Hill radius. This factor of $1/3$ arises when it is taken into account the rotation of the clump, which makes it more prone to be disrupted. Boley A. C. and T. Hayfield (private communication), indeed, show that a typical clump formed in GI simulations is a fast rotator, and that the rotation affects how tightly bound it is by this factor. A similar factor has been found in the work of \cite{zhu2012}, where the clumps' radii are close to half the Hill radii. We choose to follow the factor $1/3$ as it arises from 3D simulations, while the second study has been performed with 2D simulations.

When tidal disruption strips away the outer part of the clump, its mass is reduced accordingly. This makes the clump smaller, which leads to a larger migration timescale (see equation \ref{eq:tau}). The clump then slows down, so that it can have enough time to contract and to get out from the tidal-interaction disruption region. The clump then can survive a tidally downsizing phase.  

\subsection{Mass accretion}

While the clump moves in the disk, it can accrete mass from the disk itself. In this work the mass accretion rate is calculated starting from equation $17$ in \cite{boley2010}. In the latter, nevertheless, the clump is in the outer part of the disk, so that when this formula is used in the inner region, where the disk gas density is larger by orders of magnitude, it leads to unphysically large mass accretion rates. In order to fix this, we assume that for every order of magnitude increase in $\Sigma$, the mass accretion rate increments only of a factor of three, as it has been described in \cite{boley2013}. The formula for mass accretion rate then becomes:
\begin{equation} \label{eq:mdot}
\dot{M} = 1 \times 10^{-7} \times 3^{\log \left[ \Sigma / \Sigma_{100} \right] } \left( \frac{M_{\mbox{cl}}}{M_J} \right)^{2/3} \left( \frac{M_{\mbox{star}}}{M_{\odot}} \right)^{-1/6} M_{\odot} / \mbox{year}
\end{equation}

with $\Sigma_{100}$ surface density at $a = 100$ au. While the clump accretes mass, it radiates away the extra gravitational energy that comes from this accretion. If the timescale for the accretion is lower than the timescale to radiate this energy away, then the accretion gets stopped from a radiation pressure coming from the clump itself. The timescale over which the radiation happens is called Kelvin Helmotz timescale, and can be evaluated as the ratio between the gravitational potential and the luminosity of the clump. If it is assumed that the clump radiates as a black body, the Kelvin Helmotz timescale is then
\begin{equation}
T_{KH} = \frac{U}{L} = \frac{GM_{\mbox{cl}}^2}{4 \pi \sigma_{SB} T^4 r^3}
\end{equation} 
with $\sigma_{SB}$ Stephan Boltzmann constant and $T$ mean temperature of the clump. If this timescale is longer then the accretion timescale $T_M = M_{\mbox{cl}} / \dot{M}$, the mass accretion is stopped from the radiation pressure, and the mass accretion has an upper limit equal to $\dot{M}_{\mbox{max}} = M_{\mbox{cl}}/ T_{KH}$. When the clump is at $0.01$ au from the star, the mass accretion is zero, as at that location the disk has been cleared from the magnetic field of the star \citep{koenigl1991}. Moreover, we assume that, even if the inner disk has not been completely cleared by the presence of gas, the competing accretion from the hosting star through its magnetic filed lines, that in this region are by definition strong, will make the accretion onto the clump negligible. On the other hand, as the clump radius is smaller than a third of the Hill radius, we assume that it will be able to resist disruption by the star.

We compare the mass accretion rates found using this criteria with those presented by \cite{nayakshin2013}, where the back reaction of clumps onto the disk due to hydrodynamics feedback is analyzed. We find that the accretion rates are similar, as would be expected as the formulas we are using have been derived from simulations where gas cooling and radiative feedback are implemented. 

When the clump accretes mass, its gravitational energy increases as well. In order to keep constant the ratio between gravitational and thermal energies, the clump radius increases accordingly. We can assume that this process happens isothermally by comparing the accretion and cooling luminosities. The accretion luminosity is given by $L_{\mbox{acc}} = GM\dot{M} / R$, and has typical values of the order of $10^{30}$ erg/s. The cooling luminosity can be evaluated as the black body luminosity emitted by the clump $L_{cool} = 4 \pi r^2 \sigma_{SB} T_{ps}^4 $ where $T_{ps} = 2.7 \times 10^5 (v_{ff}/100)$ K is the post shock temperature and $v_{ff}$ gas free fall velocity in km/s. Typical values of the cooling luminosity are in the range $10^{41}$ erg/s. Such a large value of the $L_{cool}$ compared to $L_{acc}$ leads to a very quick lost of the heat generated by the accretion shock. As the clump temperature stays constant during this process, the radius rescales with the following rule:
\begin{equation}
r' = r + \delta r = r \left( 1 + \frac{\delta M}{M_{\mbox{cl}}}  \right)
\end{equation}

\subsection{Gap opening} \label{sec:gap}

While the clump is migrating in the disk, it can open a gap if it is massive enough that the gravitational torques that it excerts on the disk overcome the local torque given from the disk viscosity $\nu$. In this work we adopt the gap opening criteria presented in \cite{crida} and \cite{kley2012}:
\begin{equation}
\frac{3}{4}\frac{H(a)}{R_h} + \frac{50}{q \Re} \leq 1
\end{equation}
with $H(a)$ local disk scale hight from vertical hydrostatic equilibrium condition $H(a) = c_s \times a / v_K$ ($v_K$ local Keplerian velocity) and $\Re = a^2 \Omega^2 / \nu = (a^2 \Omega) / (\alpha H(a)^2)$ Reynolds number ($\omega$ Keplerian angular velocity at $a$, $\alpha$ viscosity parameter). Although this result has been obtained for viscous disks, we adopt it also in a self-gravitating disk case.

When a gap is opened, the migration timescale becomes similar to the local viscous diffusion timescale \citep{lin1986}:
\begin{equation}
\tau_{\mbox{vis}} = \frac{a^2}{\nu}
\end{equation}
For a typical case of gap opening at $50$ au, the local viscous diffusion timescale is between $10^5$ yrs and $10^6$ yrs, depending on the assumed viscosity value. 

The mass accretion onto the clump is affected as well from the gap opening process. The local $\Sigma$ decreases by a factor of 10 (cfr figure 1 in \cite{crida}). We therefore assume that the mass accretion is the same you would have if the clump were embedded in a disk with this lower density surface. This is not completely correct, though, as it assumes that mass accretion would proceed in the same way even though the clump is not actually embedded in the disk gas now. Therefore, we run a second set of simulations, where mass accretion is completely stopped when the gap is opened. These two extreme cases give a lower and a higher limit on the final mass of the clump.

\section{Results} \label{sec:res}

We create a set of $1000$ clumps built with IC and $1000$ clumps built with ICM and evolve each of them in four different scenarios: A,B,C and C\_m0. Clumps collapse, migration and tidal disruption are implemented in all these four configurations. The other physical mechanisms are implemented in different ways between the sets, in order to separate the effects of each of them on the clump evolution. Table \ref{tab:sim} highlights the differences between the scenarios.

As there is no general consensus in the community about the value for the viscosity parameter $\alpha$, scenario C is run for two different spatially and temporally uniform values: $\alpha = 0.05$ and $\alpha = 0.005$. This is considered to be a realistic range from  simulations and model studies. Indeed, the results of circumstellar evolution models presented in  \cite{vorobyov2009a,vorobyov2009b} show that values of $\alpha \geq 0.1$ lead to the destruction of the disk in less then 1 Myr, incompatible with the observational data. Moreover, they show that values of $\alpha \geq 0.01$ manage to reproduce the mass accretion rate of gas onto young disks but decreases the chances to have GI. The work of \cite{lodato2004} shows that when the disk fragments, $\alpha = 0.05$. We expect the value of $\alpha$ in the disk region inside the fragmenting region to be lower than this, as otherwise the disk would have experienced fragmentation al lower semimajor axis. Therefore, we take this value as an upper limit. Smaller values of $\alpha$ have been derived in studies which try to enligth the physical processes that generate turbulence; the study \cite{nelson2004}, for example, where turbulence is generated via ideal MHD, gives a value $\alpha = 7 \times 10^{-3}$. Out choice $\alpha = 0.05$ and $\alpha = 0.005$ is therefore meant to explore all the range of possibilities. In the first case, the clumps are never able to open a gap; therefore the results from simulations C with $\alpha = 0.05$ are the same as simulation B, and they are not shown. In the second implementation ($\alpha = 0.005$), mass accretion onto the clump when the gap opening criteria is fullfilled is implemented in the two limiting cases described in section \ref{sec:gap}. See table \ref{tab:sim} for the details of the sets of simulations performed.

\begin{figure}
  \centering
    \includegraphics[scale=0.15]{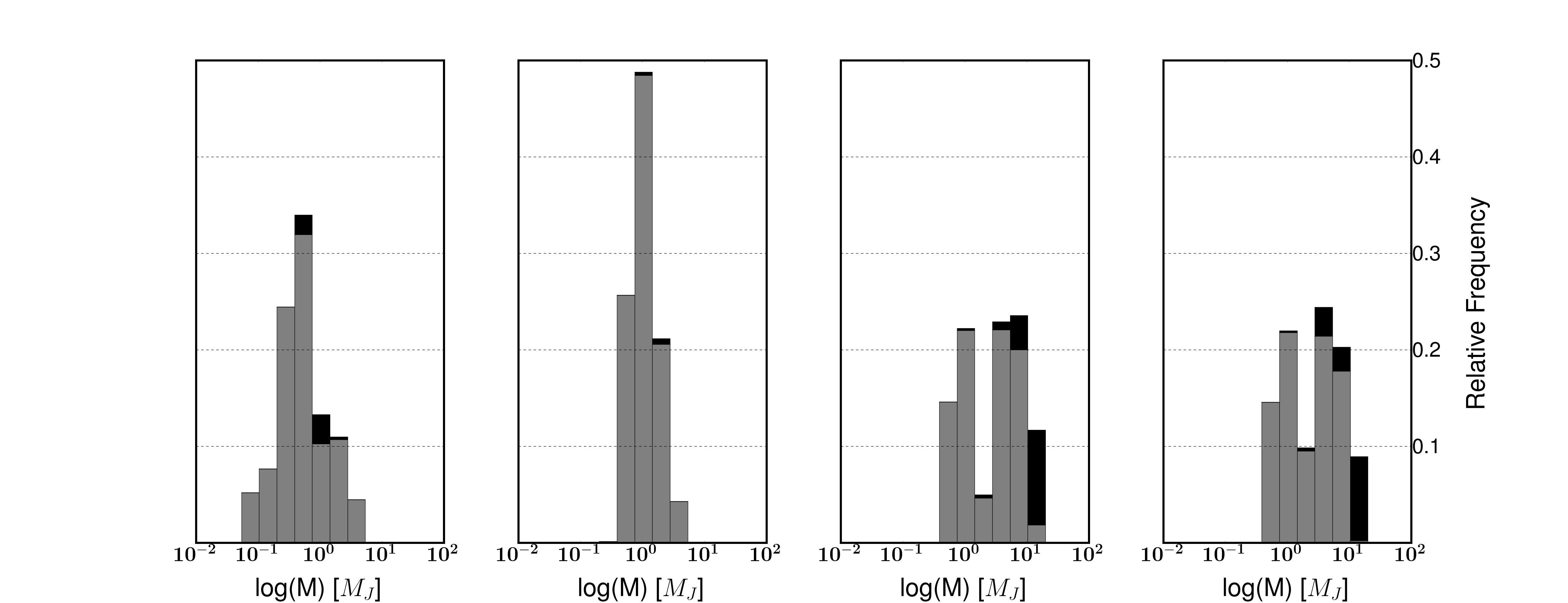}
    \caption[Final Mass]{\textbf{Final Mass} Histograms of the final mass distribution for surviving clumps (in $M_J$ units). From left to right: scenario A,B,C and C\_m0. In black, clumps that survive after tidal downsizing. In grey, clumps than never undergo tidal downsizing.}\label{tab:mass}.
\end{figure}

\begin{figure}
  \centering
    \includegraphics[scale=0.15]{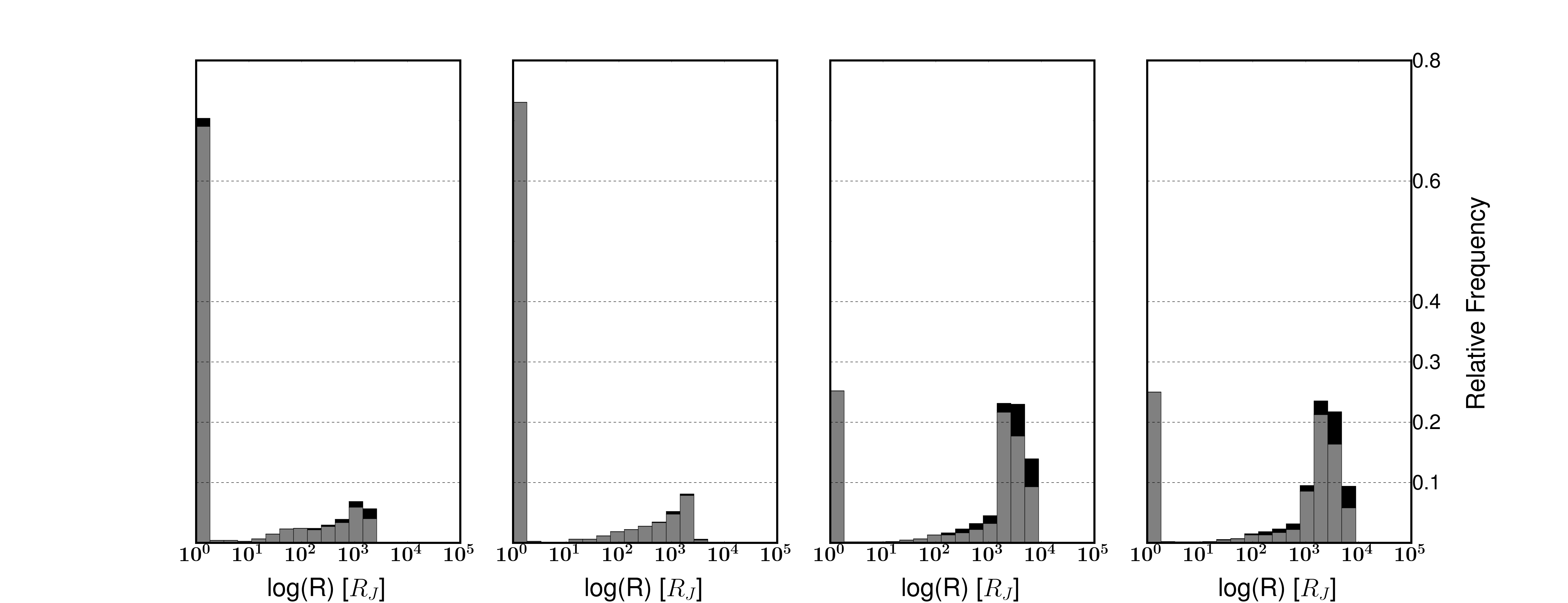}
    \caption[Final Radius]{\textbf{Final Radius} Histograms of the final radius distribution for surviving clumps (in $R_J$ units). From left to right: scenario A,B,C and C\_m0. In black, clumps that survive after tidal downsizing. In grey, clumps than never undergo tidal downsizing.}\label{tab:rad}.
\end{figure}

\begin{figure}
  \centering
    \includegraphics[scale=0.15]{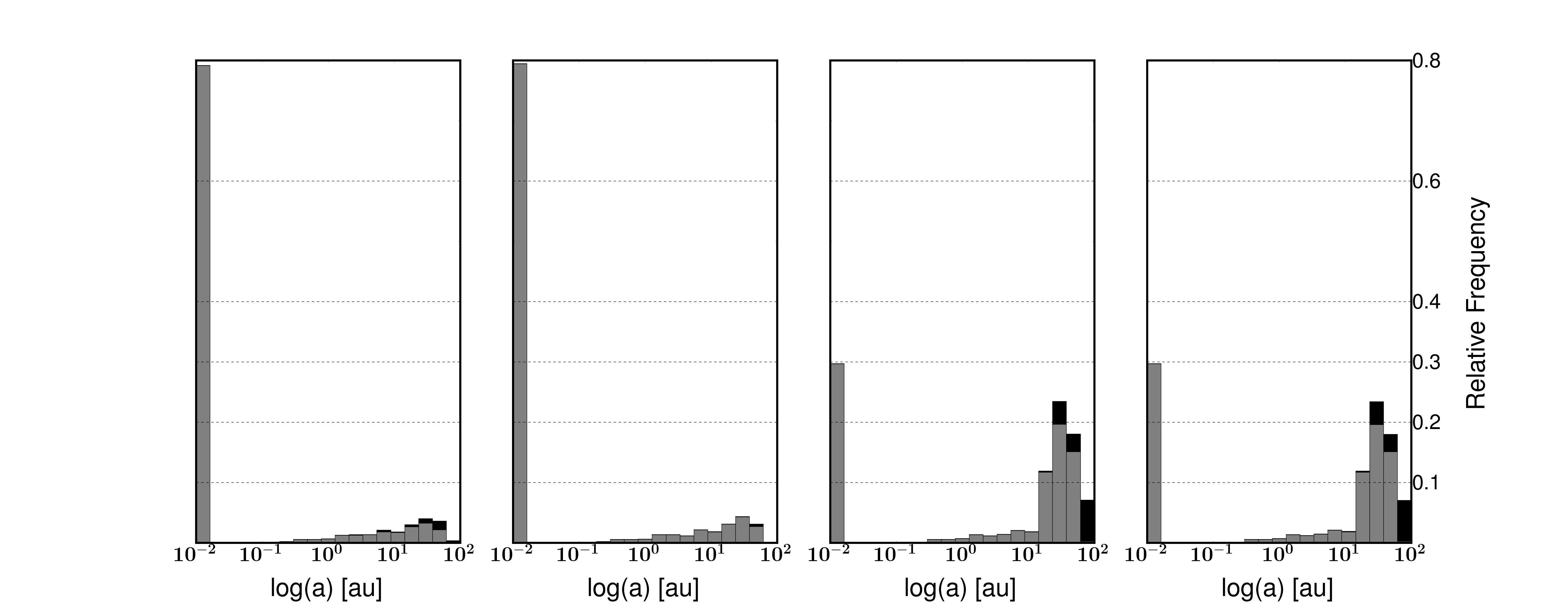}
    \caption[Final Semimajor Axis]{\textbf{Final Semimajor Axis} Histograms of the final semimajor axis distribution for surviving clumps (in au units). From left to right: scenario A,B,C and C\_m0. In black, clumps that survive after tidal downsizing. In grey, clumps than never undergo tidal downsizing.}\label{tab:a}.
\end{figure}

\begin{figure}
  \centering
    \includegraphics[scale=0.22]{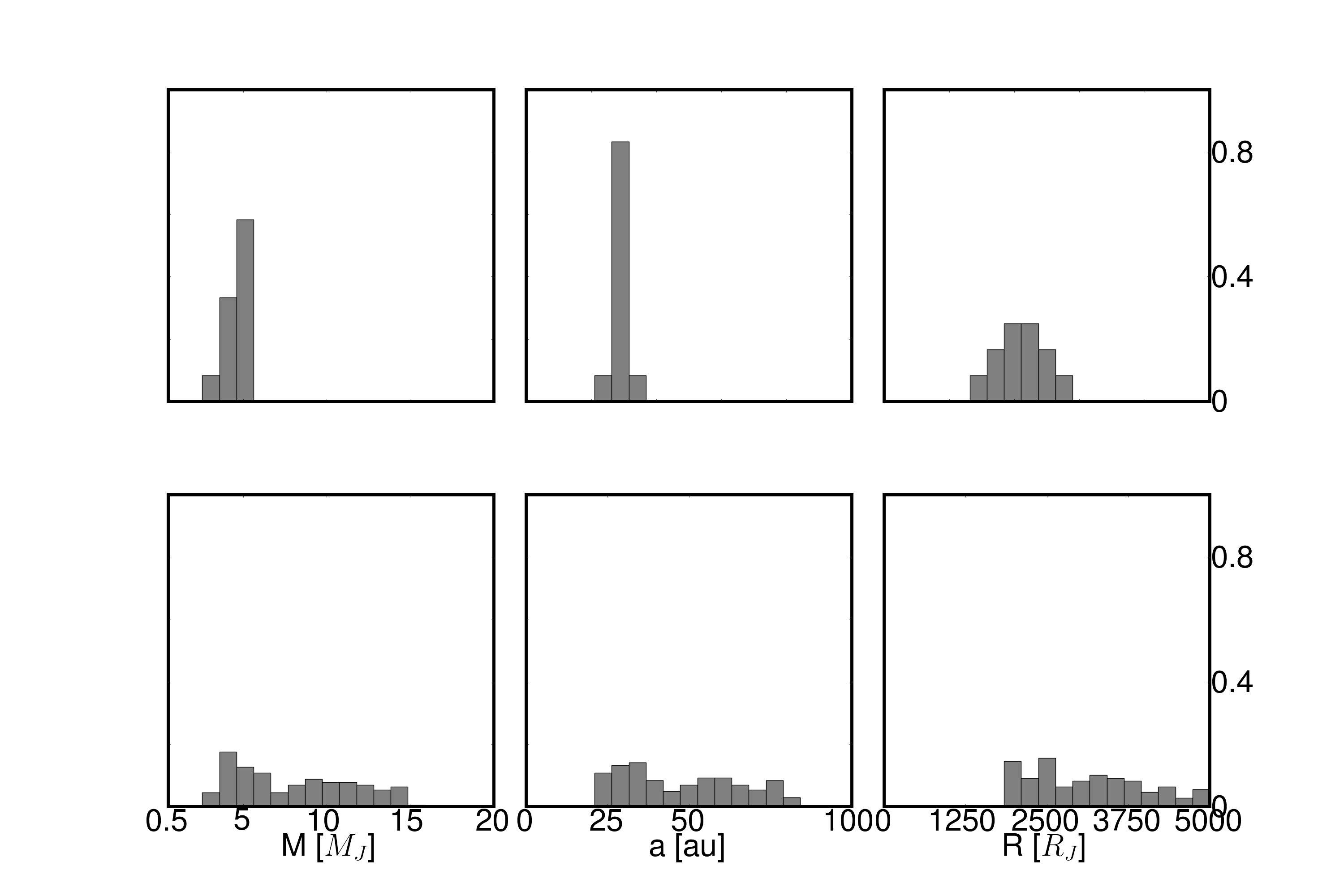}
    \caption[Gap Opening]{\textbf{Gap Opening} Histograms of (from left to right) the mass (in $M_J$ units), semimajor axis (in au units) and clump radius (in $R_J$ units) for the clumps when they open a gap in the disk in scenario C with $\alpha = 0.005$. On the top: initial condition IC. On the bottom: initial condition ICM}\label{tab:gap}.
\end{figure}

\begin{figure}
  \centering
    \includegraphics[scale=0.2]{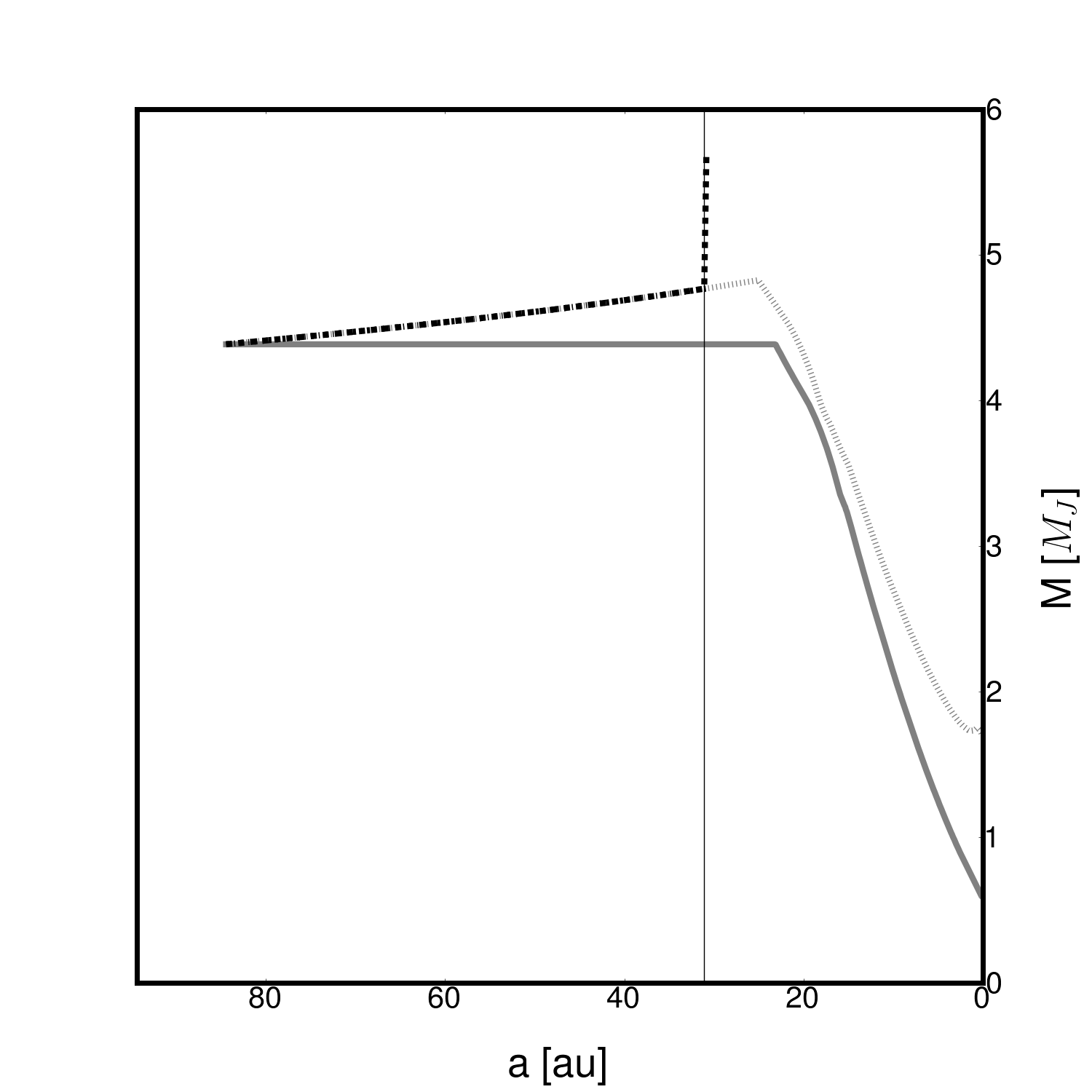}
    \includegraphics[scale=0.2]{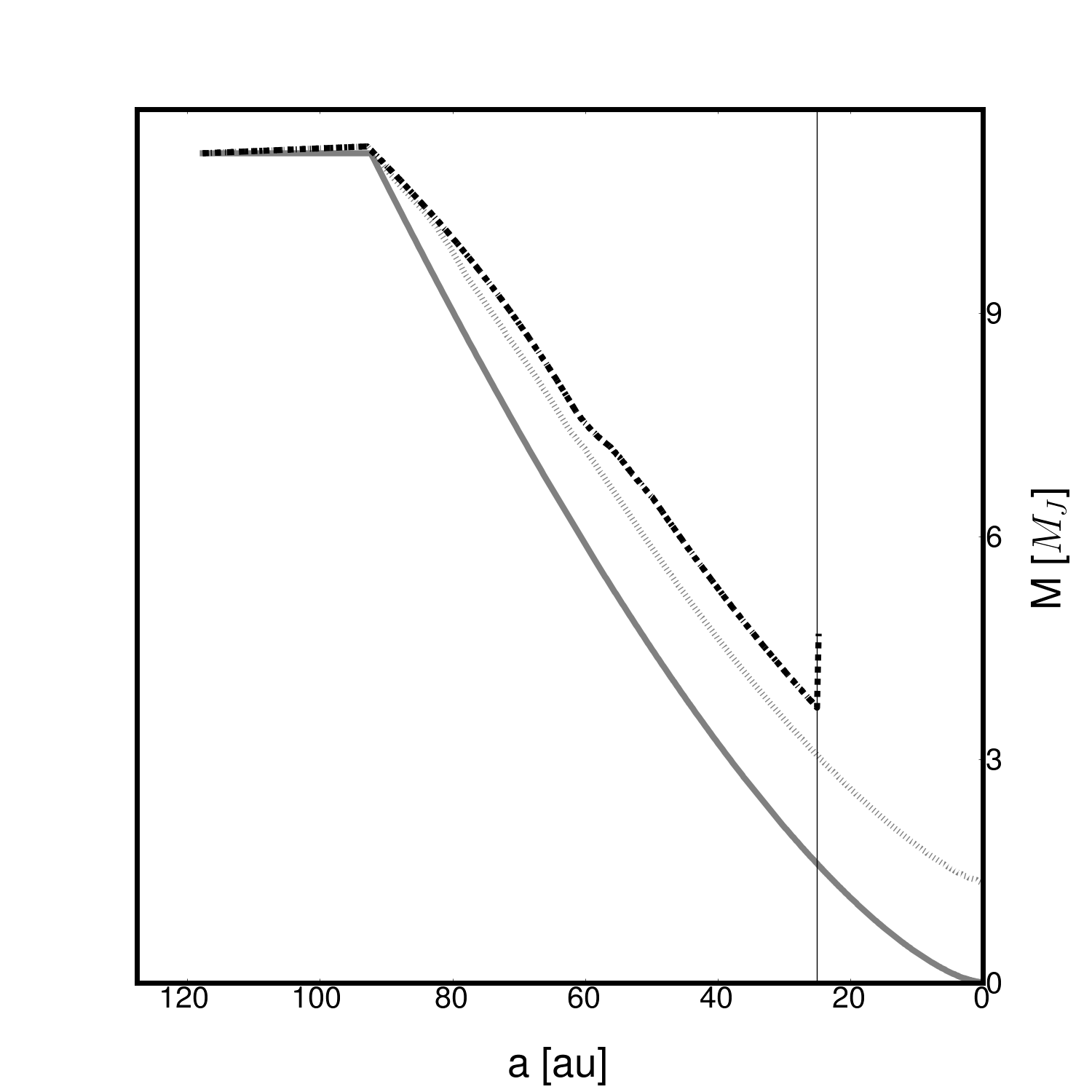}
    \caption[Clump Evolution]{\textbf{Clump Evolution} Evolution of the clump mass as a function of semimajor axis. On the left, one clump from the initial condition set IC, on the right one clump from the initial condition set ICM. The lines represent scenario A (solid gray line), scenario B (dashed gray line) and scenario C with $\alpha = 0.005$ with (dashed black line) and without (dashed black line) mass accretion during gap opening. Units are Jupiter masses and au.}\label{tab:clp}.
\end{figure}

Table \ref{tab:prob} shows the probabilities for the different outcomes of the clump evolution in the simulations sets: clump survival without undergoing tidal disruption, clump survival after being affected by tidal downsizing and clump disruption. It is also shown the probability of gap opening in scenario C. Figures from \ref{tab:mass} to \ref{tab:a} show the distribution of final mass, radius and semimajor axis for the survived clumps. Figure \ref{tab:gap} shows mass, radius and semi-major axis position of the clumps that are able to open a gap in scenario C. Figure \ref{tab:clp} shows the evolution of two clumps (one from the set IC, one from ICM) in the four different scenarios. The latter figure stresses how the inclusion of the gap opening criteria, in the case of a low viscous disk, is the physical mechanism that dominates the final outcome of the clump evolution. It is remarkable that in figure \ref{tab:clp} the same clump can either survive or be disrupted depending on whether or not there is gap opening.

We run the models until the protoplanet has reached the second collapse phase, so this is the time that we indicate as "final" from now on. From that point onward the clump will collapse dynamically to a very small size, becoming virtually insensitive to tidal effects, but can in principle continue to accrete mass and will continue to migrate. However, to date there are no 3D numerical simulations exploring this late phase of the protoplanetary collapse and with enough resolution to resolve the clump collapse and the interaction with the disk and the hosting star at the same time. Therefore, it is not possible to construct a simple model. Furthermore, disk evolution, driven by accretion onto the star and irradiation/photoevaporation, should be accounted if we had to probe longer timescales. Therefore we decided to postpone the study of the late phase to a future work. However, in the Discussion session we comment on the results of trial runs in which we have continued to evolve the protoplanets while neglecting disk evolution.

The survival rate is larger then $50$ \% in all cases. Survival of clumps without undergoing tidal downsizing is rare, and happens mainly when the gap opening is implemented. In this case, moreover, the surviving probability becomes larger, reaching $90$ \% in the case with massive initial conditions.

The final mass distribution depends both on the initial conditions chosen and the number of mechanisms that we have taken into account. When the gap opening is not implemented, indeed (scenario A and B) only small clumps manage to survive, as the larger ones have migration timescales very short and get disrupted. Clumps have masses up to $3 M_J$ (with typical value of $1.5 M_J$) for the initial condition IC and $5 M_J$ (with typical value of $3 M_J$) for initial condition ICM (with few exceptions). Once the gap opening is added, though, even more massive clumps can survive. Indeed gap opening is more luckily to happen for massive clumps, and once the gap has been open the clump will be able to survive. This leads to a mass distribution spread up to $15 M_J$ in the case with initial condition ICM, when the probability of a clump opening a gap is quite high (see table \ref{tab:prob}). Once the gap has been open, mass accretion is still possible in scenario C, while it has been stopped in scenario C\_m0. These two different implementation do not significantly affect the final mass of the clumps, as the gap is opened in the external part of the disk, where anyway mass accretion onto the clump is not significant.

The final distribution of the clump radius shows that only clumps that are able to collapse to very small objects can survive. Moreover, more massive clumps (ICM) need to collapse more then the lighter ones (IC) to survive, as their Hill radius is shorter. The inclusion of gap opening prevents the disruption of some of the large clumps, by preventing them from getting too close to the star. The final radius distribution shows that the typical final radius is of the order of tens of Jupiter radii, with the largest values are found for those clumps which are able to open a gap, therefore never experiencing tidal disruption, and can be has high as $10^4 R_J$. This result is confident with what has been observed in \cite{helled2011}: in this work only clumps in isolation were studied, which are comparable to our case of clumps that open a gap. The final radius that has been found is of the order $10^3-10^4 R_J$ also in their work. This is a non surprising result, as SCC and the successive slow contraction still have to happen, and those processes are supposed to reduce the radius by order of magnitudes.  

The final semimajor axis distribution shows that survived clumps tend to sit very close to the star: almost all of them, indeed, have a final semimajor axis inside 1 au from the star. When the initial clump is lighter (IC) it can sit also at larger distances (up to 75 au), while massive clumps always get very close to the star. This is due to the very short migration timescale for massive clumps. The inclusion of gap opening helps the massive clumps to be retained at larger distances, between 25 and 75 au. Still, a significant fraction of the survived clumps (15 \%) sits in a very inner orbit.

Figure \ref{tab:gap} shows the mass, radius and semimajor axis of the clumps when they open a gap in scenario C for both the sets of initial conditions. The clumps that are able to open a gap are the more massive once. The semimajor axis is between 25 an 70 au, making it possible the formation of proto giant planets at those distances from the star.

\end{multicols}

\begin{table}
\caption{Survival probability without tidal interaction (TD), survival probability after tidal interaction, disruption and gap-opening probability for the clumps in the eight different sets of simulation.}
\begin{center}
\begin{tabular}{|l||c|c|c|c|||c|c|c|c|}
\hline
 Set & & IC & & & & & ICM & \\
 \hline
  & Surv & TD & Disr & Gap Open. & Surv & TD & Disr & Gap Open. \\
 \hline
 $A$ & $4$ \% & $53$ \% & $43$ \%  & - & $0$ \% & $10$ \% & $90$ \%  & - \\
 $B$ & $0$ \% & $57$ \% & $43$ \%  & - & $0$ \% & $30$ \% & $70$ \%  & - \\
 $C$ & $6$ \% & $51$ \% & $44$ \%  & $5$ \% & $16$ \% & $72$ \% & $12$ \%  & $79$ \% \\
 $C \_ \mbox{m} 0$ & $6$ \% & $51$ \% & $44$ \%  & $5$ \% & $16$ \% & $72$ \% & $12$ \%  & $79$ \% \\
\hline
\end{tabular}
\end{center}
\label{tab:prob}
\end{table}

\begin{multicols}{2}

\section{Discussion} \label{sec:dis}

The results presented in this work represent a first step towards a more detailed model of the formation and evolution of clumps via GI. Despite the simple physics implemented, we can already conclude that GI appears to be a luckily mechanism to form gas giant planets. From the results presented, indeed, it is possible to extrapolate some general behaviour:

\begin{itemize}

\item the survival rate for the light clumps is never lower then $50$ \%. In the case with massive initial conditions, the survival rate is very low (the miminum being 10 \%) as long as gap opening is not taken into account. Once this is implemented, indeed, survival rates are even higher than in the low mass case, as a more massive clump is more prone to open a gap and therefore survive (up to 88 \% of surviving rate). This means that the probability of a clump formed via GI to become a proto-gas giant planet is quite high. Not only: the semi-major axis and mass distribution of the surviving clumps shows that these proto-planets are in the position and have the right mass to be considered in most of the cases the progenitors of Hot Jupiters. 

\item The inclusion of gap opening affects the final fate of clumps only if the disk viscosity is low, $\alpha = 0.005$. Larger values, $\alpha = 0.05$, indeed, do not affect the clump evolution. If this is the case, we observe that the gap is always opened between 25 and 70 au. Once the gap is opened, the clump migrates on a viscous timescale. Moreover, gap opening is easier for massive clumps, which are those more subject to being tidally destroyed. Combining these two factors together, the net result of including gap opening in a low viscosity disk is to create a population of survived clumps with high masses and further out in the disk.

\item The simpler scenario A for light clumps and the gap opening scenarios (for both initial conditions) shows that there is a (small) population of clumps which survives without ever undergoing a phase of tidal disruption.
In the first case this happens as light clumps have large migration timescales, so they are able to reach SCC before they get close to the star, while in the second case the gap prevents some clumps from getting in the inner part of the disk.

\item None of the clumps undergoes tidal downsizing twice. If they survive the first downsizing, they will become small enough to have the contraction timescale winning over the migration one.

\end{itemize}

The study presented in this paper aims at giving well physically motivated estimates for the evolution of clumps formed via GI, using the latest results from simulations and other works. Despite this, in most cases we had to include only a first order approximation of the physical mechanisms involved, partially for simplicity and partially because a complete description of some of those mechanisms is still missing. As an example, the herein derived contracting timescales are supposed to be a better estimates compared to the ones currently available, which derive from 1D collapses \citep{helled2006,forgan}. Once 3D asymmetries and angular momentum transport inside the clump itself are taken into account, indeed, the contraction is quicker than usually calculated. Nevertheless, improvements on those estimates are expected through a more detailed description of the physics involved in the collapsing phase. In particular, the inclusion of flux limited diffusion in the cooling routine is a natural improvement which is supposed to slow down the contraction. A detailed study of the inclusion of this effect (Galvagni et al. in preparation) nevertheless shows that this time increase is within one order of magnitude; we can therefore use the currently available timescale as a first approximation study. Moreover, there are some physical mechanisms which haven't been taken into account and which could lead to a shortening of the contracting timescale. One case is the opacity evolution due to the chemical changes in the dust composition during the collapse, which have been shown to shorten the contracting timescale \citep{helled2011}. The contracting timescale herein used can therefore be considered conservative.

The description of tidal downsizing is done in a first order approximation, neglecting effects due to the reaction of the clump to the tidal field. Nevertheless, the use of $R_h/3$ instead of the more common $R_h$ as the maximum radius the clump can have before being tidally disrupted should take care of these second order effects, leading our work to confident results. A more precise study of the dependence of the reducing factor of the Hill radius on the characteristics of the clump and of the local disk is currently under investigation. A more precise treatment of the tidal downsizing phase would impact mainly the clump radius evolution. Indeed taking into account the heating which comes from the tides between the star and the clump would increase the clump radius. The final radius distribution for the survived clumps is therefore not completely trustable.

The clumps have been studied in isolation, as if each of them formed alone in the disk from the fragmentation of spiral modes induced from GI. From simulation studies \citep{boss2011,vorobyov2013} and from the existence of multi-companion systems that could have formed via GI \citep{marois2010}, it appears that formation of two or more clumps at the same time is not rare. The study of the more massive set of initial condition ICM is meant to partially cover this scenario. Despite this, our study neglects the clump-clump interaction (in the same way as core-core interaction has been neglected in CA synthesis population studies, \cite{alibert2005,mordasini2012} ), which could play a major role in determining the final population of survived clumps. It has to be noted, though, that in most simulations which show the formation of only two clumps, the spiral fragments at the opposite extremes, so that the reciprocal interaction of the clumps can be considered a second order effect.

One other effect that has being neglected in our study is photoevaporation from the star onto the clumps \citep{nayakshin2012} once they reach the inner part of the disk. Moreover, the possibility that the presence of clumps at the inner edge of the disk can lead to FU Ori outburst has been neglected as well \citep{armitage2001,zhu2011}. From our study, it appears that at that stage clumps are already dense enougth to be able to survive these phenomena, at least partially; nevertheless a consistent investigation of these effects needs to be included.

The inclusion of magnetic fields could also play a major role, as it is known that MRI influences the disk viscosity and surface density profile \citep{mohanty2013}, which dominate the interaction between the clump and the disk. A detailed study of the dependence of the clump evolution on the disk structure is still needed.

The migration considered in this work is always inward migration. Nevertheless, it is known \citep{bate2003,baruteau} that migration of clumps is usually not smooth in one direction, and could even be outward. \cite{nayakshin2012} relates the migration time to the mass evolution of the clump, with particular focus on the tidal downsizing phase. According to their findings, it is possible to have outward migration driven by mass loss. In the description given in this work of the tidal downsizing process, though, the mass loss rate is quite slow (of the order of a few Jupiter masses per thousand years), so that the change in angular momentum due to this process is not able to significantly affect the migration rate. 

In this study core formation has been completely neglected, therefore we can not make any quantitative prediction on the formation of rocky/icy planets through tidal downsizing. A comparison between the time that it takes for the clumps evolved in this study to be subjected to tidal downsizing and the temperature evolution of the core given in \cite{forgan} would show that, if the cores that form at the center of a clump were able to survive the tidal downsizing phase, they would for the large part be in the rocky state. Indeed the inner temperature would have overcome the critical value of $130$ K far before tidal downsizing starts, due to the fast collapsing timescales assumed. A secondary effect of these fast timescales would be that the cores would be smaller that usually calculated \citep{forgan}, as they have less time available to growth inside the pressure maxima of the clump.

In the recent work by \cite{forgan} a population of clumps formed via GI is followed with a semi analytical model for $10^6$ years. The main result of this study is the formation of a large population of Brown Dwarf objects at large distance from the central star ($40-60$ au). Despite the impossibility of a direct comparison between this result and the analysis presented in this paper, as we stop at SCC, it is possible to say that our results are not consistent, as the survived clumps from our model are too small and close-in in the disk to make it reasonable that future evolution can lead to the same conclusion. This disagreement lies in the different implementations of the physical mechanisms in the two models. One major difference, is the assumed mass and semimajor axis of the IC. \cite{forgan}, indeed, assumes as initial mass the Toomre mass and an initial semimajor axis between 20 and 60 au, while we have the initial mass decreased by one order of magnitude from the Toomre value, following the 3D numerical results of \cite{boley2010}, and a more distant initial semimajor axis. Moreover, the contraction timescale in \cite{forgan} comes from analytical estimates, and it is larger by orders of magnitude than the timescale we use, which comes from high resolution numerical simulations. The last major difference between those two works is that the gap opening criteria adopted is not the same, and as we know gap opening plays an important role in the evolution of the clumps. In particular, since the clumps from \cite{forgan} are more massive, they are much more efficient in opening gaps at large distances, which may lower the tidal mass loss compared to our simulations. This biases further the final mass distribution to higher values.

\cite{zhu2012} presents the study of the evolution of a set of clumps formed via GI until they encounter the inner disk boundary. Despite the fact that the population of clumps that have been simulated in this work is too few to make a statistics (less than 20), it is still remarkable that the survival rate and the rate of clumps that open a gas is comparable with our findings. The main differences are that in the work of \cite{zhu2012} those clumps have not been followed in the very inner part of the disk, so we can't make any direct comparison with the clumps we observe surviving at $0.01$ au from the star. Moreover, the typical mass of clumps is larger than the one observed in this work, close to the Brawn Dwarf range. This peculiarity could be given by the fact that the simulations performed in \cite{zhu2012} are 2D simulations, and therefore they are not ideal for following the formation of fragments in GI and their collapse, as those processes are inherently three dimensional \citep{boley2010}. Moreover, \cite{mayer2008} shows that the resolution needed in grid codes, like the one used in the work \cite{zhu2012}, in order to correctly simulate the fragmentation and collapsing processes of the clumps, is larger than the one implemented in their work. Using a low resolution would on one hand prevent the formation of small clumps, and on the other hand artificially stop the contraction, leading to an overestimate of mass and radius of the typical clumps which could explain the discrepancy between our results. 

Finally, as we have explained above, we stop running our models at the beginning of the phase of second core collapse, which 
starts no later than $10^5$ yr after the initial time. Therefore, strictly we only make predictions for the properties of
a population of protoplanetary clumps in the pre-dissociation phase.  While tidal effects will cease to be important once
the clump collapses dynamically to planetary sizes and densities, migration and accretion can still be important. We have
run a subset of the clumps forward in time, finding that, irrespective of whether or not they are allowed to open a gap,
in about a million year all the protoplanets end up at the inner boundary. This is nothing else than the "fast migration"
problem encountered in population-synthesis models for planets forming via core-accetion \citep{alibert2005,mordasini2012,mordasini2013}.

Therefore, in disk instability, as well as in core-accretion, one needs to invoke a suppression of migration
or some stochastic effects that lead to a variable direction of migration, with some clumps moving outward rather than
inward. That the latter can happen is suggested by numerical simulations of fragmenting disks indeed in which multiple
clumps form \citep{durisen2007,boley2010,boss2011}. Additionally, if the gas disk is rapidly dissipated, dynamical
scattering of protoplanets can lead to fast rearrangement of their orbits, accompanied by mass segregation \citep{papaloizou2001}.

\section{Conclusions} \label{sec:con}

The study herein presented aims at understanding the fate of clumps formed via GI in circumstellar disks. In order to do that, we studied the evolution of a set of clumps, one per time, coupling their contraction with the interaction with the disk and the central star. We performed four set of simulations, in order to add step by step mass accretion and gap opening (for both a low and a high viscosity disk).

Our results show that a large fraction of the clumps survives, contrary to previous claims appeared in the literature \citep{zhu2012,forgan}. 

The higher surviving fraction is due to the fast collapse timescales in the dissociation phase. Most importatly, such a fast
collapse timescale is estimated for the first time based on the results of 3D hydro collapse simulations \citep[][and Galvagni et al. in prep.]{galvagni}. Furthermore, most of the clumps formed via GI could in principle be precursors of Hot Jupiters. 

Taken face-value, the chance that they are the progenitors of massive gas giant planets at distances between 20 and 75 au from the star is not negligible as well. However, a naive extension of our models beyond the pre-disspciation phase would lead to the prediction that all clumps have to become Hot Jupiters or be engulfed by their host star since migration timescales are always shorter than 1 Myr, no matter whether or not
gap opening takes place. This reflects the same problem found in core-accretion, namely that migration has to be much slower or be somehow stochastic, with inward-directed migration being only one of the possibilities, in order to be consistent with the wide range of semi-major axis found among the eoplanet population. Alternatively, disk dissipation has to be faster than the migration time so that protoplanets can stop migrating sooner and undergo gravitational scattering that redistributes their orbits.

The physical mechanism that seems to play a major role in shaping the properties of the population of clumps until the pre-dissociation phase is gap opening. The efficiency of gap opening is strongly tied with disk viscosity in turn. For a low viscosity disk the survival probability for the clumps get close to $50$ \% for low mass clumps and to $90$ \% for massive ones. In reality disk viscosity will be spatially and time dependent, likely transitioning from a high viscosity state soon after fragmentation has taken place (when the disk is still unstable and therefore gravitoturbulent) to a lower viscosity state in which gap opening will be effective. Models incorporating disk evolution and a more realistic prescription for viscosity will have to be investigated in the future.

In summary, our results show that GI can in principle produce protoplanetary clumps that survive the first evolution phase, up to SCC, when they are more prone to be disrupted. Therefore those objects could in principle become a large fraction of the population of present-day gas giants, including Hot Jupiters, as much as CA. An important caveat, however, is the interaction gas-solid, which has been neglected in the present study. It is therefore not possible to make a prediction of the actual structure of the planet, and a clear connection with known planets and their measured density is beyond the aim of this work. 

A problem associated with excessively migration occurs here as it does in CA. Indeed, while clumps form at much larger radii than cores in the core-accretion model, and hence have a larger distance to cover before they reach the inner disk, they also have more time to do so since GI is expected to happen early in disk evolution. A possible mechanism to stop the inward migration, which has been suggested also in CA studies, is photoevaporation of the disk \citep{alexander2012,alexander2013}. Future studies will have to elucidate the role of migration of clumps in long-term simulations in order to construct a more realistic model for the orbital evolution of clumps relative to what we have done here.

\section{Acknowledgments}
The authors thank A. C. Boley for useful discussions and suggestions and the anonimous referee whose comments improved the quality of the paper. MG thanks F. Heitsch for hosting her during the development of this work and for interesting hints.

\end{multicols}

\newpage

\newpage

\def\mnras{MNRAS}
\def\icarus{Icarus}
\def\apj{ApJ}
\def\aap{A \& A}
\def\apjl{ApJL}
\def\pasp{PASP}
\def\nat{Nature}

\newpage
\clearpage

\bibliographystyle{mn2e}
\bibliography{bibliomia}

\end{document}